# Inconsistencies of the Highly-Cited-Publications Indicator

Michael Schreiber

*Institute of Physics, Chemnitz University of Technology, 09107 Chemnitz, Germany.*

*E-mail: schreiber@physik.tu-chemnitz.de*

One way of evaluating individual scientists is the determination of the number of highly cited publications, where the threshold is given by a large reference set. It is shown that this indicator behaves in a counterintuitive way, leading to inconsistencies in the ranking of different scientists.

**Introduction**

The h-index, which was introduced by Hirsch (2005) has been criticized by Waltman and van Eck (2009, 2012) and Bouyssou and Marchant (2011) because it can lead to inconsistencies in the way in which scientists are ranked. This means that for the purpose for measuring the overall scientific impact of a scientist the h-index exhibits counterintuitive behavior. Waltman and van Eck (2012) discuss a large family of bibliometric indicators as alternatives to the h-index, but they pay special attention to the highly-cited-publications (HCPs) indicator. It is the purpose of the present investigation to show that the HCPs indicator suffers from similar inconsistencies.

To avoid misunderstandings, I note that "the word 'consistency' has different meanings in different disciplines or even different areas of the same discipline" (Ye, 2012). In the present context the discussed inconsistencies refer to counterintuitive behavior of the h-index and the HCPs indicator in the ranking of individual scientists which occur due to performance improvement measured in terms of the numbers of papers or citations, respectively.

In the following, I present three situations, in which the HCPs indicator behaves in a counterintuitive way, because the same performance improvement changes the ranking of individual scientists, or because the aggregation of data changes the ranking. It is assumed that the performance is measured in reference to a large dataset where the threshold of 10 citations defines which publications are counted as highly cited. The reference set should be large enough, so that the small changes in the example cases do not influence this threshold.

**Inconsistency with respect to relative performance improvement**

In agreement with Waltman and van Eck (2012) the following property is demanded:

> *If two scientists achieve the same relative performance improvement, their ranking relative to each other should remain unchanged.*

In order to demonstrate that this property is violated by the HCPs indicator, let us consider two scientists with two publications only, as listed in Table 1: In the original situation O scientist V has one publication with 10 citations and one uncited paper, i.e., one HCP. Scientist W has two publications with 5 citations each, i.e, the same total number of papers and citations as V, but no HCPs. Consequently according to the HCPs indicator in this evaluation V has shown a better performance than W.

Waltman and van Eck (2012) discuss a relative performance improvement in terms of the number of papers. In contrast I shall consider a performance improvements in terms of citations. Let us suppose the following relative performance improvement R (see Table 1): All papers receive more citations, namely an increase of 100%, i.e., their citation counts are doubled. As a result, V now has one paper with 20 citations and still one uncited paper, which still means one HCP. On the other hand W has now two papers with 10 citations each, i.e., two HCPs. Thus the performance of W looks better than the performance of V. Their ranking has reversed, although they have both achieved exactly the same relative performance improvement.

A somewhat more fancy example is based on a slightly more elaborate citation distribution for 3 scientists X, Y, Z, presented in Tables 2 and 3, where the symbols E, G, M, B, P distinguish between excellent, good, medium, bad, poor numbers of 15, 12, 9, 6, 3 citations, respectively, in the original situation O as denoted in Table 2. In Table 3 the citation distributions which are assumed in this example for X, Y, and Z are given in terms of these groups: Scientist X has 4 publications with good, medium, and bad performance each. In comparison, for scientist Y two of the G papers are replaced by one E and one M paper, and two B papers by one M and one P paper. The same change is repeated to create the citation distribution of Z. As a consequence, the three scientists have the same total numbers of 12 papers with 108 citations.

In the original situation O, only E and G papers are above the HCP threshold of 10 citations. Therefore the scientists X, Y, and Z achieve a score of 4, 3, 2 HCPs, respectively, see Table 3. So in the ranking X is better than Y and Y is better than Z.

Let us now consider a relative performance improvement of 33%, i.e., all citation numbers are increased by one third. This is denoted as case R1 in Table 2 and the result is given in Table 3: Now E, G, and M papers are above the threshold of 10 citations. Thus X, Y, Z score 8, 9, 10 HCPs, respectively, so that their ranking is inverted: Z is now better than Y, and Y is better than X.

A further relative performance improvement by one forth compared to the situation R1 (or equivalently by two thirds compared to the original situation O) as shown in Table 2, raises also the B papers to or above the threshold of 10 citations. It thus yields a score of 12, 11, 10 HCPs for scientists X, Y, Z, respectively, and therefore again reverses the ranking, see Table 3: Now X is once more better than Y, and Y is better than Z. In my view this is a very strange behaviour and I conclude that the HCPs indicator leads to inconsistencies in the ranking when scientists achieve the same relative performance improvement.

The h-index suffers from the same problem as shown in Table 4. For example, if a scientist P has published 4 papers, three with 3 citations each and one uncited, the h-index is $h_P = 3$ and thus better than that of scientist Q who is assumed to have published one paper with 3 citations and three papers with two citations each, so that $h_Q = 2$. Again as above we start with the same total number of papers and citations for P and Q. The relative performance improvement by an increase of 100% in the citation counts yields $h_P = 3$ and $h_Q = 4$, i.e. a reversed ranking.

**Inconsistency with respect to absolute performance improvement**

Now I intend to show that the HCPs indicator violates the following consistency property which was also proposed by Waltman and van Eck (2012):

> *If two scientists achieve the same absolute performance improvement, their ranking relative to each other should remain unchanged.*

The previous examples have been constructed in such a way, that they can also be used in the present discussion. Marchant (2009) and Bouyssou and Marchant (2011) have discussed the same property and called it independence. Like Waltman and van Eck (2012) they consider a performance improvement by adding further publications, while I shall keep the number of papers fixed and analyze a performance improvement in terms of more citations.

I first consider scientists V and W again and assume that all their papers attracted five more citations each. This absolute performance improvement A is listed in Table 1, too. Consequently, V now has one paper with 15 citations and one paper with 5 citations, which means one HCP. On the other hand, W has now two papers with 10 citations, i.e., two HCPs and therefore W is ranked better than V. The same absolute performance improvement of the citation records of V and W has changed the ranking as compared to the original situation, where V ranked better than W.

For the 3 scientists X, Y, and Z discussed already in the second example above I assume an absolute performance improvement of 3 citations for every publication, denoted in Table 2 as the case A1. As shown in Table 3, the score of HCPs is the same as for the situation R1 and thus the ranking in the case A1 is reversed in comparison with the original situation O, just as it was reversed for the case R1. Again a further performance improvement by the same amount, i.e., 3 more citations for each publication (denoted in Table 2 as case A2) yields the same scores as the case R2 and thus the same ranking, which means once more an inversion of the ranks in comparison with the case A1 as also indicated in Table 3. Again the conclusion is, that the HCPs indicator provided inconsistent rankings of X, Y, and Z.

As Table 4 shows, the h-index suffers from the same problem, in this case the same absolute performance improvement of the above discussed citation records of P and Q by 2 citations to each publication reverses the ranking.

**Inconsistency with respect to aggregation**

In order to show the inconsistency of the HCPs indicator for rankings at different levels of aggregation, I consider the following property:

> *If scientist X is ranked higher than scientist Y in each of two different time intervals, then X should be ranked higher than Y also when both time spans are combined.*

This is similar to the property which was called consistency by Bouyssou and Marchant (2011) for the accumulation of citation records of various scientists into datasets for departments, which means combining sets of papers. In contrast, like in the two previous sections, I consider the combination of the citation records of the same scientists in different periods of time, which means a summation of the citations for the same papers.

In order to show how this requirement can be violated by the HCPs indicator, in Table 5 I start with the very simple example of the two scientists V and W again. Now I assume that the original citation record O has been achieved in the first time period which could be one year for example. I assume that in the second time interval, like the second year, both scientists do not write any paper, but receive the same number of citations to their previous publications as in the first time span. So in both time intervals V has one HCP, while W has no HCPs. Thus V is ranked higher than W in both time periods. However, if both time intervals are combined, the aggregated citation distributions coincide with the situation R in Table 1: They show one publication with 20 citations and still one uncited paper for V, and two papers with 10 citations each for W. This means that V has still one HCP only, while W achieved two HCPs. Consequently the ranking is inverted by the accumulation of the citations over the longer time span: W turns out to be ranked higher than V.

Also for this consistency property I have constructed a more elaborate example, presented in Table 6. Here, the scientists S, T, U have four papers each, which received a total of 20 citations in the first year as well as in the second year. For S, the citations are equally distributed, each paper got 5 citations in each year. For T, one paper is and remains uncited, the 5 citations are additionally given to the first paper in the first year and to the second paper in the second year. For U, there are two uncited papers, and the other two papers received 10 citations each in each year. These citation counts are given in Table 6, as well as the number of HCPs, namely 0, 1, 2 for S, T, U, respectively, in both years. Thus U is better than T and T is better than S in both years. However, if the citation counts are accumulated for both years, then we have 4, 3, 2 HCPs for S, T, U, respectively, as shown in Table 6. Thus the ranking is again inverted: From the aggregated citation distributions we conclude that S is better than T and T is better than U.

For the h-index the same inconsistency with respect to aggregation can be demonstrated. For this purpose I return to the very simple example of two scientists P and Q again. Using the same argumentation as for V and W in Table 5, I assume that the original citation record O has been achieved in the first time span and the performance improvement is achieved in the second time interval, where all papers are assumed to receive the same number of citations as in the first time span, but no new papers have to be taken into account. Thus P is better than Q in terms of the h-index considering both time spans separately. However, if the citation counts are accumulated, the ranking is reversed, see Table 7. I admit that the present example is based on an unusual evaluation of the citation record, because in the determination of the h-index one usually does not distinguish different citation periods, but rather looks at the evolution of the index with increasing length of the citation period, or distinguishes different publication intervals.

**Concluding remarks**

It has been shown that the HCPs indicator behaves in a counterintuitive way in specific situations, which can be interpreted as an inconsistent behavior leading to inconsistencies in the ranking of scientists, which achieve the same performance improvements. This is similar to the observations which have been described by Waltman and van Eck (2012) with respect to the h-index. The first two consistency properties in that analysis are the same as in the present work. However, in that investigation the performance improvement was achieved by additional publications. In the present study, the number of publications is kept constant, and the performance improvement is obtained by increasing the number of citations.

In the same spirit, the third consistency requirement in the present paper is based on the aggregation of citation records of the same scientists, but for different time intervals, so that again the same publications can be considered. In contrast, Waltman and van Eck (2012) discuss the aggregation of the citation distributions of different scientists.

In any case, the present results show that the HCPs indicator suffers from similar difficulties as the h-index. In my view, demanding the consistency with respect to an improvement of the citation records is as reasonable as demanding it with regard to the increase of the number of papers at least for the relative improvement. Of course this depends on the particular assumptions on which one bases the concept of scientific impact or scholarly influence, as discussed by Ravallion and Wagstaff (2011). In particular for an absolute improvement of the citation record, the situation is not so clear-cut. There are good reasons to assume a diminishing marginal influence which means that the first citation to a given publication is considered to have the highest influence and the impact of each citation is decreasing with increasing number of citations to the same paper (Ravallion and Wagstaff, 2011). From the receiving point of view, "the gain in influence indicated by the first citation received by a publication that was previously ignored must surely by larger than the gain in influence that is implied by an extra citation received by the most cited, and hence most influential, paper in a discipline" (Ravallion and Wagstaff, 2011, p. 326). From the citing side one could argue that citations are sometimes the result of a preferential attachment process, because it is more likely that somebody refers to a publication that has already been frequently cited. In other words: Not all citations are created equal. In this case an inversed ranking due to an absolute performance improvement would not appear to be unreasonable.

Waltman and van Eck (2012) favored the HCPs indicator, because they did not take possible inconsistencies in terms of improved citation records into consideration. The present investigation thus questions the conclusion that the HCPs indicator is a better choice; it remains a matter of taste whether the HCPs indicator is considered to be more attractive or not. As shown above, the consistency with respect to more citations is violated by the h-index and the HCPs indicator. The consistency with regard to additional publications is violated by the h-index but not by the HCPs indicator. This leads me to the assessment that the HCPs indicator is not as bad as the h-index.

In the above analysis I have constructed very simple examples. In these examples the inconsistency with respect to aggregation looks very similar to the inconsistencies observed with respect to the performance improvement in the previous sections. But these are not equivalent. The condition that scientist X is ranked higher than scientist Y can be fulfilled by very different citation records, therefore the consistency property with respect to aggregation can be violated more easily. Thus this property imposes a very strong requirement.

The examples provided above comprise a very small number of publications with specific citation counts. But it is easy to see that the results are not changed in any way, if the same number of HCPs is added to the citation distribution of each scientist, as well as the addition of an arbitrary number of lowly cited papers does not alter anything, as long as they remain lowly cited during the performance improvements or the accumulation procedure.

Ye (2012) argued that the inconsistency of the h-index is not an issue in dynamical systems. The same criticism could be applied to the inconsistencies with respect to the HCPs indicator discussed in the present investigation. However, I do not believe that the objection is valid in either case. Ye argued that small changes do not lead to inconsistencies and that large changes need time. This is true, but the described counterintuitive behaviour of the HCPs indicator as well as of the h-index remains counterintuitive even when it needs a long time to occur. Of course the presented simple examples will not be found in real dynamical systems, but this does not mean that in real systems similar situations could never be observed which show the same unreasonable effect, namely that the same performance improvement in the citation distributions of two scientists over a certain time span can reverse the ranking of two scientists, and thus leads to counterintuitive behaviour.

**Acknowledgement**

I thank L. Waldmann for helpful comments.

Table 1. Citation record of two scientists (V, W) with two publications each; giving the number of citations (cits.) and the number of HCPs for the original situation (O), after a relative improvement of 100% (R), and after an absolute improvement of 5 citations for each paper (A); numbers of citations in bold face indicate HCPs.

|   | Original O | | Rel. imp. R | | Abs. imp. A | |
|---|---|---|---|---|---|---|
|   | Cits. | HCPs | Cits. | HCPs | Cits. | HCPs |
| V | **10**+0 | 1 | **20**+0 | 1 | **15**+5 | 1 |
| W | 5+5 | 0 | **10**+**10** | 2 | **10**+**10** | 2 |

Table 2. Number of citations for excellent (E), good (G), medium (M), bad (B), and poor (P) papers for the original situation (O), after a relative improvement of 33% (R1) and 67% (R2), as well as an absolute improvement of 3 citations per paper (A1) and 6 citations per paper (A2); numbers of citations in bold face indicate HCPs.

| Case | Change | E | G | M | B | P |
|---|---|---|---|---|---|---|
| O | ±0 | **15** | **12** | 9 | 6 | 3 |
| R1 | 33% | **20** | **16** | 12 | 8 | 4 |
| R2 | 67% | **25** | **20** | **15** | 10 | 5 |
| A1 | 3 | **18** | **15** | **12** | 9 | 6 |
| A2 | 6 | **21** | **18** | **15** | **12** | 9 |

Table 3. Citation distributions in terms of the groups E, G, M, B, P from Table 2 for three scientists (X, Y, Z); the resulting HCPs score for the numbers of citations in the cases O, R1, R2, A1, A2 as given in Table 2, and the corresponding ranking of the three scientists.

|   | Citation distribution | | | | | HCPs score | | | Rank | | |
|---|---|---|---|---|---|---|---|---|---|---|---|
| Scientist | E | G | M | B | P | O | R1/A1 | R2/A2 | O | R1/A1 | R2/A2 |
| X | 0 | 4 | 4 | 4 | 0 | 4 | 8 | 12 | 1 | 3 | 1 |
| Y | 1 | 2 | 6 | 2 | 1 | 3 | 9 | 11 | 2 | 2 | 2 |
| Z | 2 | 0 | 8 | 0 | 2 | 2 | 10 | 10 | 3 | 1 | 3 |

Table 4. Citation record of two scientists (P, Q) with four publications each; giving the number of citations (cits.) and their h-indices for the original situation (O), after a relative improvement of 100% (R), and after an absolute improvement of 2 citations for each paper (A); numbers of citations in bold face indicate papers which count for the h-index.

|   | Original O | | Rel. imp. R | | Abs. imp. A | |
|---|---|---|---|---|---|---|
|   | Cits. | h | Cits. | h | Cits. | h |
| P | **3**+**3**+**3**+0 | 3 | **6**+**6**+**6**+0 | 3 | **5**+**5**+**5**+2 | 3 |
| Q | **3**+**2**+2+2 | 2 | **6**+**4**+**4**+**4** | 4 | **5**+**4**+**4**+**4** | 4 |

Table 5. Same as Table 1, but for different time spans; numbers of citations in bold face indicate HCPs.

|   | First year | | Second year | | Both years | |
|---|---|---|---|---|---|---|
|   | Cits. | HCPs | Cits. | HCPs | Cits. | HCPs |
| V | **10**+0 | 1 | **10**+0 | 1 | **20**+0 | 1 |
| W | 5+5 | 0 | 5+5 | 0 | **10**+**10** | 2 |

Table 6. Same as Table 5, but for three scientists with different citation records; numbers of citations in bold face indicate HCPs.

| Scientist | First year | | Second year | | Both years | |
|---|---|---|---|---|---|---|
|   | Cits. | HCPs | Cits. | HCPs | Cits. | HCPs |
| S | 5+5+5+5 | 0 | 5+5+5+5 | 0 | **10**+**10**+**10**+**10** | 4 |
| T | **10**+5+5+0 | 1 | 5+**10**+5+0 | 1 | **15**+**15**+**10**+0 | 3 |
| U | **10**+**10**+0+0 | 2 | **10**+**10**+0+0 | 2 | **20**+**20**+0+0 | 2 |

Table 7. Same as Table 4, but for different time spans.

|   | First year | | Second year | | Both years | |
|---|---|---|---|---|---|---|
|   | Cits. | HCPs | Cits. | HCPs | Cits. | HCPs |
| P | **3**+**3**+**3**+0 | 3 | **3**+**3**+**3**+0 | 3 | **6**+**6**+**6**+0 | 3 |
| Q | **3**+**2**+2+2 | 2 | **3**+**2**+2+2 | 2 | **6**+**4**+**4**+**4** | 4 |